\documentclass[12pt,tightenlines,
    onecolumn,
    preprintnumbers,
    amsmath,
    amssymb,
    preprint,
    nofootinbib,
    showpacs,
]{revtex4}
\usepackage{color}
\usepackage{graphicx}%
\usepackage{amsmath}
\usepackage{amssymb}
\usepackage{bm}
\usepackage{enumerate}

\newcommand{\nn}{\nonumber}

\begin{document}
\title{ Geometrical properties of the trans-spherical solutions in higher dimensions}

\author{Gungwon Kang}
\email{gwkang@kisti.re.kr}
\affiliation{Korea Institute of Science
and Technology Information (KISTI), 335 Gwahak-ro, Yuseong-gu,
Daejeon 305-806, Korea }
\author{Hyeong-Chan Kim}
\email{hckim@cjnu.ac.kr}
\affiliation{Division of Liberal Arts, Chungju National University, Chungju, 380-702, Republic of Korea
}
\date{\today}%
\author{Jungjai Lee}
\email{jjlee@daejin.ac.kr ; CA}
\affiliation{ Department of Physics, Daejin University,
Pocheon, 487-711, Korea.
}%


\bigskip
\begin{abstract}
\bigskip

We investigate the geometrical properties of static vacuum $p$-brane solutions of Einstein gravity in $D=n+p+3$ dimensions, which have spherical symmetry of $S^{n+1}$ orthogonal to the $p$-directions and are invariant under the translation along them.
The solutions are characterized by mass density and $p$ number of tension densities.
The causal structure of the higher dimensional solutions is essentially the same as that of the five dimensional ones.
Namely,  a naked singularity appears for most solutions except for the Schwarzschild black $p$-brane and the Kaluza-Klein bubble.
We show that some important geometric properties such as the area of $S^{n+1}$  and the total spatial volume are characterized only by the three parameters such as the mass density, the sum of tension densities and the sum of tension density squares rather than individual tension densities.
These geometric properties are analyzed in detail in this parameter space and are compared with those of 5-dimensional case.

\end{abstract}
\maketitle

\section{Introduction}

Recently, the physical meaning of the two parameters in static
vacuum hypercylindrical solutions in five dimensions~\cite{Kramer}
was correctly interpreted in Ref.~\cite{CHLee}. In the analogy of
weak field solutions for a cylindrical matter source distributed
along the fifth direction uniformly, the author identified the two
parameters with ``mass" and ``tension" densities~\footnote{They are
indeed gravitational mass and tension densities rigorously defined
at the asymptotic infinity~\cite{tra}.}. It is also pointed out that
the well-known Schwarzschild black string solution corresponds to
the case that the tension-to-mass ratio is exactly one half. This
means that the Schwarzschild black string, which was believed to be
a family of solutions characterized by the mass density only, is
indeed a special case of a wider class of solutions characterized by
tension density as well. Note also that in $5$-dimensional
spacetime, there is another class of stationary solutions which is
characterized by the ``mass" and ``momentum" densities along the
extra-directions~\cite{jlkim,kimjl}. The Schwarzschild black string
background is known to be unstable under small gravitational
perturbations along the fifth direction - the so-called
Gregory-Laflamme (GL) instability~\cite{GL,after}. Therefore,
studying the physical role of tension might give a better
understanding about what really causes the GL instability.

In Ref.~\cite{kang} the geometric properties of this class of
spacetimes with arbitrary tension in five dimensions were
investigated in detail. Some of main properties are as follows.

\begin{itemize}
\item The solutions are classified by the tension-to-mass ratio $a$.
The event horizon exists only when the tension density is a half of
the mass density, {\it i.e.} $a=1/2$. Only in this case the
spacetime can be called a black string. All spacetimes having other
values of the tension-to-mass ratio than $1/2$ and $2$ have a naked
singularity at `center'.

\item

Even though there is a naked singularity instead of an event horizon,
light radiated from it is infinitely red-shifted provided that $a<2$.

\item The geometry of some subspaces behaves interestingly. For $a <
1/2$ or $a > 2$ the area of $S^2$ sphere monotonically decreases
down to zero as one approaches to the naked singularity from
infinity as usual. For $1/2 < a < 2$, however, it bounces up and
increases again up to infinity at the naked singularity as in the
geometry of a wormhole spacetime.

\item On the other hand, the proper length of a segment along the fifth
direction shrinks down to zero for $a > 1/2$, but expands to
infinity for $a < 1/2$ as one approaches the naked singularity.

\item Although the $S^2$ area and the segment length compete each other,
the total area of the segment $S^2 \times L$ turns out to decrease
monotonically down to zero at the singularity except for $a = 1/2$.
For the case of $a=1/2$ the area of $S^2$ becomes finite at the
horizon and the scale factor of the fifth direction is constant.
\end{itemize}
The geodesic motions in this spacetime were also studied in Ref.~\cite{Gwak}.

In this paper we investigate how much these features change if the spacetime dimensionality becomes higher than five.
For the case of $ a < 1/2$, for instance, since the area of $S^2$ sphere shrinks to zero whereas the proper length along the fifth direction diverges, one may expect that the total area might not
shrink down to zero if one increases the number of extra dimensions large enough.
In addition, since there will be more number of tension parameters as the number of extra dimensions increases, one may wonder if there are more black brane solutions in addition to higher dimensional Schwarzschild black branes.

Higher dimensional spacetime solutions having translational symmetry along extra dimensions and spherical symmetry on slices perpendicular to them have been discovered by many authors in the literature in different contexts~\cite{Myers,Agnese,Bronnikov:1995uh,Bro2,stability,Ohta1}.
For instance, much more general solutions even in the presence of dilatonic scalar fields and antisymmetric forms were
found in Refs.~\cite{Bronnikov:1995uh,Bro2,Ohta1}. (See also references therein.)
However, the full and detailed analysis of the geometry and correct interpretation of the `tension' parameter have not been done so far as long as we know.

In section II, we study the geometric properties of the
trans-spherical vacuum solutions in $D=n+p+3$ dimensions in detail. In section III, we analyze the
causal structure of the solution. Finally in section IV, we
summarize our results and discuss physical implications of them.

\section{Geometric Properties of Solutions}

We consider static solutions for the vacuum Einstein equations in
$D=n+p+3$ dimensions, which are invariant under translations along
the extra $p$-dimensions and are spherically symmetric on the
$(n+2)$-dimensions transverse to the $p$-dimensions. The most
general form of the metric with the symmetries may be written as
\begin{eqnarray} \label{metric:ansatz}
ds^2 &=&g_{\mu\nu}dx^\mu dx^\nu=-H_0(\rho)dt^2
+G(\rho)\left(d\rho^2+\rho^2d\Omega_{(n+1)}^2\right)+
   \sum_{i=1}^p H_i(\rho) dz_i^2 .
\end{eqnarray}
Here $H_0$, $G$, and $H_i$ are functions of the isotropic coordinate
$\rho$ only. This class of spacetime solutions will be characterized
by the $p$-number of tension densities $\tau_i$ along $z_i$ and the
ADM mass density $M$ associated with the spatial and time
translation symmetries, respectively~\cite{tra}. Interestingly, it
turns out that the exponents in the metric component functions
depend only on the dimensionless tension-to-mass ratio $a_i$ defined
as
\begin{eqnarray}
\tau_i = a_i M.
\end{eqnarray}
For convenience, we define the sums of $a_i$ and $a_i^2$ by
\begin{eqnarray} \label{a}
 a = \sum_{i=1}^p a_i, \quad  \bar a^2= \sum_{i=1}^p a_i^2 ,
\end{eqnarray}
where $i$ runs over spatial extra-dimensions. Note that these
definitions for  $a$ and $\bar a$ restrict values of $\bar a$ such
that $\bar a \geq |a/\sqrt{p}|$ as shown by the shaded region in
Fig.~1 below.
The solutions in this general higher dimensions were found by
several authors~\cite{Bronnikov:1995uh,Bro2,Ohta1} and can be
expressed as
\begin{eqnarray} \label{Sol2}
H_0(\rho)&=&\left|\frac{1-m/\rho^n}{1 +
m/\rho^n}\right|^{\sqrt{\frac{n+1}{n}}\frac{\frac{2(p+n-a)}{p+n+1}}{\sqrt{
     \bar a^2+1- \frac{(a+1)^2}{p+n+1}}}  }, \\
G(\rho)&=&\left(1+\frac{m}{\rho^n}\right)^{
            \frac{4}n} \,\left|\frac{1-m/\rho^n}{1 +
 m/\rho^n}\right|^{\frac{2}{n} \left( 1-\frac{a+1}{p+n+1}\frac{\sqrt{n(n+1)}}{\sqrt{\bar
    a^2+1 -\frac{(a+1)^2}{p+n+1} }}  \right)}
     , \nn \\
H_i(\rho)&=& \left|\frac{1-m/\rho^n}{1 +
m/\rho^n}\right|^{\sqrt{\frac{n+1}{n}}\frac{2\left( a_i -\frac{a+1}{p+n+1}\right) }{\sqrt{
     \bar a^2+1- \frac{(a+1)^2}{p+n+1}}}  } \nn
\end{eqnarray}
by using the ADM mass and ADM tensions. Here the integration
constant $m$ is related to the ADM mass $M$ by
\begin{eqnarray}
m=\left[\frac{n}{n+1}\left(
     \bar a^2+1- \frac{(a+1)^2}{p+n+1}\right)
    \right]^{\frac1{2}}\frac{4\pi G_D M}{\Omega_{(n+2)}}, \label{m:Ma}
\end{eqnarray}
where $\Omega_q=q\pi^{q/2}/\Gamma(q/2+1)$, and $G_D$ is the
$D$-dimensional Newton constant. Note that the quantity inside the
square root in Eq.~(\ref{m:Ma}) is positive definite for all real
values of $a_i$. The solutions~(\ref{Sol2}) are described by $p+1$
independent parameters, $m$ and $a_i$. It is interesting to see that
the functions $H_0$, $G$, and $\prod_{i\geq 1} H_i$ depend only on
the sums of $a_i$ and $a_i^2$ ({\it i.e.}, $a$ and $\bar a^2$), not
on the individual values of $a_i$. The double wick rotations $t\to
iz_i$ and $z_i \to i t$ of the metric~(\ref{metric:ansatz}) also
satisfies the vacuum Einstein equation.

Let us analyze geometrical properties of the spacetime solutions
described above and see how they depend on the spacetime
dimensionality.
First of all, one sees that these spacetimes become flat as either
$\rho \rightarrow \infty$ or $\rho \rightarrow 0$.
The same happens in 4D Schwarzschild and Reissner-Nordstrom solutions written in isotropic coordinates.
From the form of solutions, we find that the component functions of the metric,
$H_0$, $G$, and $H_i$ show different behaviors as $\rho \to K$,
depending on the sign of their exponents. Here
\begin{eqnarray}\label{K:m}
K= |m|^{1/n}.
\end{eqnarray}
Note that the region with $0<\rho<K$ is another copy of the
spacetime region covered by $K < \rho < \infty$ as in the case of
the $5$-dimensional one~\cite{kang}~\footnote{This can be easily
seen by the coordinate transformation, $\rho \rightarrow
K^2/\rho$.}. Although arbitrary values of $m$ ({\it e.g.}, $M$) are
allowed, we assume non-negative values of ADM mass density for
discussions below. Thus, $m = K^n$.

Note also that the case of uniform tensions with $a_i= 1/(n+1)$  for all $i$ corresponds
to the well-known Schwarzschild black $p$-brane in higher dimensions,
\begin{eqnarray}\label{Schwarz}
ds^2= - \left(\frac{1-K^n/\rho^n}{1+K^n/\rho^n}\right)^2 dt^2 + G(d\rho^2+\rho^2 d\Omega_{(n+1)}) +
    \sum_{j=1}^p dz_j^2 ,
\end{eqnarray}
where the event horizon is located at $\rho=K$. The other well known
special cases are given by ($a_i=n+1,~~a_{j\neq i}=1$), the
Kaluza-Klein bubble solution,
\begin{eqnarray}\label{bubble1}
ds^2= -dt^2 + G(d\rho^2+\rho^2 d\Omega_{(n+1)}) +\left(\frac{1-K^n/\rho^n}{1+K^n/\rho^n}\right)^2 dz_i^2+
    \sum_{j\neq i}^p dz_j^2 ,
\end{eqnarray}
which are related with the Schwarzschild black $p$-brane solution
mentioned above through the double Wick rotation. Notice also that
the metric becomes singular at $\rho=K$. It turns out that this is a
coordinates singularity for the case of the Schwarzschild black
$p$-brane and the Kaluza-Klein bubble~\footnote{The conical
singularity in the case of Kaluza-Klein bubble is removed by
compactifying the $z_i$ coordinate, {\it e.g.}, $z_i\sim
z_i+(2^{2+2/n}\pi K)/\sqrt{n} $.}. Otherwise, it is a curvature
singularity as will be shown below in detail.

By investigating the spatial geometries of several spacelike
hypersurfaces, we classify the solutions in the parameter space of
$a_i$. Note first that, in $4+1$ dimensions with $n=1=p$, the area
of $S^2$ diverges if $1/2 < a< 2$, finite if $a=1/2$ or $2$, whereas
it vanishes if $a>2$ or $a<1/2$~\cite{kang}.
In general, the area of the $S^{n+1}$ sphere at the $z_1, \cdots, z_p=\mbox{constant}$ surface is given by
\begin{eqnarray} \label{A}
A_{n+1}(\rho) &=& \Omega_{(n+2)}(\sqrt{G}\rho )^{n+1}\\
   &=&\Omega_{(n+2)}
   \rho^{n+1}\left(1-\frac{K^n}{\rho^n}\right)^{\frac{n+1}{n}\left(1
    -\alpha\right)} \times
    \left(1+\frac{K^n}{\rho^n}\right)^{\frac{n+1}{n}\left(1
    +\alpha\right)}. \nn
\end{eqnarray}
Here $\alpha$ is defined as
\begin{eqnarray} \label{coef}
&&\alpha = \frac{a+1}{p+n+1}\frac{\sqrt{n(n+1)}}{\sqrt{\bar
    a^2+1 -\frac{(a+1)^2}{p+n+1} }}  .
\end{eqnarray}
As $\rho \to \infty$, this area increases as usual. However, as
$\rho \to K$, one can see that the property is crucially dependent
on the signature of the exponent $(1-\alpha)$. The area of $\rho=K$
surface has a non-vanishing finite value if $1-\alpha=0$. As shown
by the thick curve in Fig.~1 this equation gives the hyperbola
defined by the following equation on $a$ and $\bar a$ space
\begin{eqnarray} \label{An:fin}
 \frac{(n+1)^2+p}{(n+p+1)^2}\,(a+1)^2- \bar a^2= 1
\end{eqnarray}
with a restriction of $a>-1$. Note that this hyperbola intersects with the line $\bar a= a/\sqrt{p}$ one time if $n\geq (p+1)/(p-1)$ and two times if $1\leq n< (p+1)/(p-1)$.
\begin{figure}[htbp]
\begin{center}
\includegraphics[width=.8\linewidth,origin=tl]{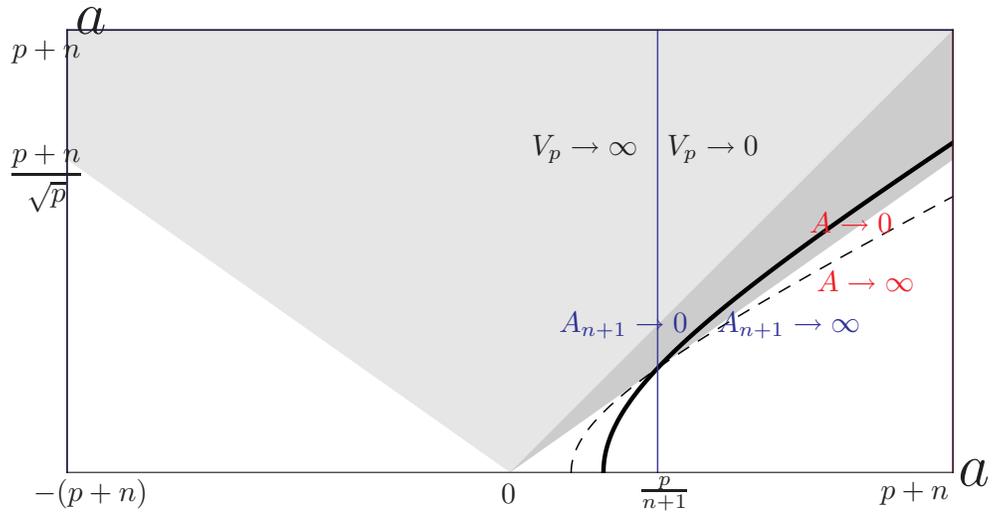}
\end{center}
\caption{(color online) Schematic phase diagram of higher dimensional $p$-brane solutions in $(a,\bar a)$.
The horizontal and vertical axes stand for $a$ and $\bar a$, respectively.
The definitions of $ a$ and $\bar a$ allow their values on the shaded region satisfying $\bar a\geq |a/\sqrt{p}|$.
The dark shaded region is for physical parameters based on the strong energy condition.
As $\rho$ goes to $K$, the area of the $(n+2)$-dimensional sphere $A_{n+1}$ becomes a non-vanishing finite value on the thick curve,
the volume $V_p$ for extra $p$-dimensions on the vertical black line,
and the total volume $A$ on the dashed curve.
 } \label{fig1}
\end{figure}
On the hyperbola, the area of the $\rho=K$ surface becomes
\begin{eqnarray} \label{..}
A_{n+1} = 4^{1+\frac1n}\Omega_{(n+2)} K^{n+1}=
    \Omega_{(n+2)}^{-1/n}
    \left(\frac{16n\pi G_D M}{n+p+1}\right)^{\frac{n+1}{n}}
    (a+1)^{\frac{n+1}{n}}.
\end{eqnarray}
Therefore we see that, while only the cases of $a=1/2$ and $2$ in $5$-dimensions give a non-vanishing finite area $A_2$, all solutions whose gravitational tensions are properly balanced in accordance to Eq.~(\ref{An:fin}) have non-vanishing finite area $A_{n+1}$.
Namely, in the parameter space of $p$-dimensional $a_i$,
this region
is given as the intersection of the $(p-1)$-dimensional sphere
$\sum_{i=1}^p a_i^2 = \bar a^2$ and a $(p-1)$-dimensional plane
$\sum_{i=1}^p a_i =a $ for each $(a,\bar a)$ on the hyperbola~(\ref{An:fin}) in the shaded region in Fig.~1.

If $1-\alpha<0$, the behavior of the area is basically the same as the case of $1/2<a<2$ in $5$-dimensions.
Namely, as $\rho$ decreases to $K$ from infinity, this area decreases only up to its minimum point $\rho=\rho_m$  given by
\begin{eqnarray}
\rho_m
 &=& \left(\frac{4\pi G_D M}{\Omega_{(n+2)}}\right)^{\frac1n}
  \left(n(a+1)+\sqrt{\frac{(n+1)^2+p}{(p+n+1)^2}(a+1)^2-(\bar a^2+1)}
  \right)^{\frac1n},
\end{eqnarray}
and then starts to increase up to infinity instead of decreasing
down monotonically. Therefore, this spatial geometry looks like a
wormhole. However, we point out that the $\rho = K$ surface can be
reached from the throat ({\it i.e.}, $\rho = \rho_m$) in a finite
affine time. For $n\geq (p+1)/(p-1)$, the value of $\rho_m$ does not
have a maximum. On the other hand, for $n< (p+1)/(p-1)$,  it has a
maximum. In Fig.~1, the corresponding parameter space is designated
by the right-down side of the thick curve in the shaded region.

On the other hand, if $1-\alpha>0$ this area monotonically decreases
down and shrinks to zero at $\rho=K$ as in the cases of $a>2$ or
$a<1/2$ in $5$-dimensions.  In Fig.~1, the corresponding parameter
space is designated by the left-up side of the thick curve in the
shaded region.

In the special case of string configurations ({\it i.e.}, $p=1$),  $a$ and $\bar a$ become same, and $A_{n+1}$ diverges as $\rho$ goes to $K$ in the region given by
\begin{eqnarray} \label{volume}
\frac{1}{n+1} < a < n+1.
\end{eqnarray}
At $\rho=K $, $A_{n+1}$ takes a non-vanishing finite value for the cases of
$a=\frac{1}{n+1}$ or $a= n+1$.  For the cases of $a<1/(n+1)$ or $a>n+1$, it vanishes.
For $n=1$ we recover the $5$-dimensional results.

Now let us consider the spatial geometry in extra $p$-dimensions.
Note first that, if $a_i=\tau_i/M=1/(n+1)$ for all $i$,
all scale factors of the extra space are constant since all $H_i=1$.
This case corresponds to the Schwarzschild black $p$-brane.
For given values of tension-to-mass ratios $\{a_1,a_2,\cdots ,a_{i-1},a_{i+1},\cdots ,a_p\}$, the $z_i$ direction becomes flat if
\begin{eqnarray}\label{ai:regular}
a_i = \frac{1+a_1+a_2+\cdots +a_{i-1}+a_{i+1}+\cdots +a_p}{p+n},
\end{eqnarray}
equivalently, $a_i=\frac{a+1}{p+n+1}$.
For other values of tension ratios, we see that $H_i$ in Eq.~(\ref{Sol2}) becomes singular as $\rho$ goes to $K$.
Namely, as $\rho$ decreases to $K$, the proper length of a unit segment with $\Delta z_i=1$ at $\rho=$ constant surface shrinks to zero if $a_i>\frac{a+1}{p+n+1}$, but it infinitely expands if $a_i<\frac{a+1}{p+n+1}$.
However, we point out that all these cases do not necessarily correspond to singular spatial geometry in extra $p$-dimensions.
Consider a Kaluza-Klein bubble solution along the $z_1$ direction, in which $H_1(\rho)$ vanishes at $\rho=K$ with $H_{i\neq1}(\rho)=1$.
We can easily check that there exist no curvature singularity in this case.
Indeed, the $\rho=K$ surface turns out to be a fixed point whose conical singularity can be removed
by compactifying the $z_i$ coordinate suitably as mentioned before.
Hence, the $z_1$ direction becomes a circle.
In general, if $p\geq 2$, one can find many solutions where some extra directions become circles in this way.
However, all those solutions contain curvature singularity at $\rho=K$ except for the case of the Kaluza-Klein bubble solutions.
To summarize, we find that the regular spatial geometry in extra $p$-dimensions happens only for the Schwarzschild black $p$-brane and Kaluza-Klein bubble solutions.

It is interesting to see how the total spatial volume $A$ for a given $\rho$ behaves as $\rho \to K$.
The  total volume for $\rho =$ constant surface per unit lengths
of $z_i$ with respect to an asymptotic observer is given by
\begin{eqnarray} \label{r}
A &=& A_{n+1} V_p
\end{eqnarray}
where  $V_p$ is the spatial volume of extra $p$-dimensions
\begin{eqnarray} \label{volume}
V_{p} =\prod_{i=1}^p\int_{z_i}^{z_i+1}\sqrt{H_i} dz_i  =
    \left( \frac{1-K^n/\rho^n}{1+K^n/\rho^n}\right)^{\sqrt{\frac{n+1}{n}}\frac{\frac{(n+1)a-p}{p+n+1} }{\sqrt{
     \bar a^2+1- \frac{(a+1)^2}{p+n+1}}}  }.
\end{eqnarray}
We see that the spatial volume $V_p$ becomes unit when $a = p/(n+1)$ which is designated by the vertical line in Fig.~1.
It is interesting to see that the total volume $V_p$ could be finite if the divergence at a certain direction can be precisely canceled out by the shrinking of some other directions.
As $\rho$ decreases to $K$, $V_p$ monotonically decreases to zero for $a>p/(n+1)$ and monotonically increases to infinity for $a<p/(n+1)$.

The  total spatial volume of $\rho=$constant section is given by
\begin{eqnarray} \label{r}
A &=&  \Omega_{(n+2)}\rho^{n+1}
\left(1+\frac{K^n}{\rho^n}\right)^{
            \frac{2(n+1)}n} \,\left|\frac{1-K^n/\rho^n}{1 +
 K^n/\rho^n}\right|^{\frac{n+1}{n} \left( 1-\frac{\sqrt{\frac{n}{n+1}}}{\sqrt{\bar
    a^2+1 -\frac{(a+1)^2}{p+n+1} }}  \right)}
\end{eqnarray}
Thus the behavior of the total volume is determined by the  exponent of $[1-K^n/\rho^n]$,
\begin{eqnarray}\label{A-exp}
&&\frac{
    (p+n+1)[(n+1)\bar a^2+1] -(n+1)(a+1)^2 }{
        n(p+n+1)\sqrt{\bar a^2+1 -\frac{(a+1)^2}{p+n+1} }\left(\sqrt{\bar
    a^2+1 -\frac{(a+1)^2}{p+n+1} }+\sqrt{\frac{n}{n+1}}\right)}
\end{eqnarray}
which vanishes on a hyperbola (the dashed curve given in Fig.1).
From the definitions of $a$ and $\bar a$, the numerator of
Eq.~(\ref{A-exp}) can be reexpressed in the sum of squared terms
given by
\begin{eqnarray}\label{positive}
(n+1)\sum_{i=1}^p \sum_{j=i+1}^p (a_i - a_j)^2 +\sum_{i=1}^p
[(n+1)a_i -1]^2.
\end{eqnarray}
The Eq.~(\ref{positive}) is non-negative for all real values of $a_i$.
It vanishes only when
\begin{eqnarray}\label{pta}
a_1= a_2 =... = a_p =\frac{1}{n+1}.
\end{eqnarray}
The total volume with this set of $a_i$ becomes
\begin{eqnarray}
A= 2^{\frac{2(n+1)}{n}}\Omega_{(n+2)} K^{n+1}.
\end{eqnarray}
This case of uniform tensions is exactly the same as the case in
which all extra $p$-directions are regular as mentioned above. In
addition, we point out that the geometry of the sphere is regular
since this case satisfies Eq.~(\ref{An:fin}) as well.
In fact, the solution with parameters in Eq.~(\ref{pta}) corresponds
to the Schwarzschild black $p$-brane~(\ref{Schwarz}) in higher
dimensions.

Since either $A_{n+1}$ or $V_{p}$ diverges depending on the values of tensions one may expect the total volume $A$ also diverges.
However, we claim that it never happens for any values of tension parameters.
This can be seen because the exponent of $[1-K^n/\rho^n]$ takes a positive value for all cases except for the case of uniform tensions so that the total volume $A$ vanishes as $\rho \to K$.
This happens because, even if the volume of the sphere diverges for certain values of parameters $a_i$,  $V_p$ shrinks more strongly for those values of parameters.
Similarly, for the case in which $V_p$ diverges, $A_{n+1}$ shrinks more strongly.
The idea of making divergent total volume $A$ by adding expanding extra-dimensions arbitrarily does not work.
 This is because as $p$ increases, both of the expanding rate in $V_p$ and shrinking in $A_{n+1}$ changes as well so that the net effect always gives vanishing total volume.

So far, we have not restricted the values of $a_i$.
However, their physical ranges may be given by some energy conditions.
Note that the strong energy condition in $5$-dimensional spacetime restricts the value of tension~\cite{kang}.
In  $d$-dimensional spacetimes, this extends as follows%
\begin{eqnarray} \label{strE}
T_{00}+\frac{1}{d-2}\left(-T_{00}+\sum^p T_{z_iz_i}\right) \geq 0
    \Longrightarrow (d-3)M \geq \sum \tau_i .
\end{eqnarray}
Although we do not know whether the gravitational tensions satisfy the same condition, we assume that the same restriction holds as in the case of matter fields.  Therefore, the physical range of $a$ may be given as,
\begin{eqnarray} \label{bound}
0 \leq a \leq d-3=p+n ,
\end{eqnarray}
where the first inequality comes from the positivity theorem for gravitational tension, $a_i\geq 0$~\cite{tra}.

Note that there exists a coordinate singularity at $\rho=K$. Let us
see whether this singularity is genuine one or not. The Kretschmann
invariant for the metric~(\ref{Sol2}) is
\begin{eqnarray} \label{curv:square}
&&R_{\mu\nu\rho\sigma}R^{\mu\nu\rho\sigma}=16n{K}^{2n}  {\rho }^{4 + 2n}
    ({\rho }^n  -K^n )^{-\frac{4}{n}( 1 + n - \beta_1 )}
    ( K^n + {\rho }^n)^{-\frac{4}{n}( 1 + n + \,\beta_1)} \nn \\
&& \quad \times \left\{
  2\left [ 1 + 4 \beta_1^2 +
    \beta_1^4 + n \left(3+(5 \beta_1^2 + 2\beta_2 )  +
    \beta_1^2 ( \beta_1^2 + 4 \beta_2 )\right) \right.\right. \nn \\
&&\quad\quad \left.\left.+
    n^2 \left( 2 + (\beta_1^2 + 3\beta_2)+ 2 \beta_1 ( \beta_1 \beta_2 + 2\beta_3 )
    \right)+  n^3 ( \beta_2 + \beta_2^2 + \beta_4 )
 \right ] {K}^{2n}{\rho }^{2n} \nn \right.\\
&&\quad\quad \left.- 4(1+n)\left( \beta_1 (1+\beta_1^2)
    +n\beta_1(1+ 2\,\beta_2)+ n^2 \beta_3 \right)
    ({K}^{3n}{\rho}^n + {K^{n}}{\rho}^{3n}) \right. \nn \\
&&\quad\quad \left. + (1 + n) (2 + n) (\beta_1^2
    + n\beta_2)({K}^{4n} + {\rho}^{4n})  \right \}
\end{eqnarray}
where $\beta_q= \sum_{i=0}^p (\alpha_i)^q$ with $q=1,2,3,4$. Here
$\alpha_i$ with $i=0, 1, \cdots, p$ is the half of the exponent of
$\frac{1-K^n/\rho^n}{1+K^n/\rho^n}$ in $H_i$. We see that the right
hand side of the first line becomes singular as $\rho \to K$ since
$1+n-\beta_1=1+n-\alpha>0$. Thus, there occurs a curvature
singularity at $\rho=K$ unless the term inside the curly-bracket
cancels this divergence. We find that such cancelation happens only
for the cases satisfying all the following conditions:
\begin{eqnarray} \label{beta:eq}
\beta_1 = \beta_2 = \beta_3= \beta_4=1.
\end{eqnarray}
By solving these algebraic equations above, we obtain either
\begin{equation} \label{calpha}
 \alpha_0=1, \quad\quad\alpha_1=\alpha_2=\cdots=\alpha_p=0,
\end{equation}
or, ($i=1,2,\cdots ,p$)
\begin{eqnarray} \label{bubble}
 \alpha_i = 1, \quad \alpha_0=\alpha_1=\alpha_2=\cdots =\alpha_{i-1}=\alpha_{i+1}=\cdots= \alpha_p=0.
\end{eqnarray}
The first case~(\ref{calpha}) corresponds to the Schwarzschild black $p$-brane.
The second case corresponds to the $p$ number of different Kaluza-Klein bubble solutions, which are related to the Schwarzschild black $p$-brane by the double-Wick rotations in Eq.~(\ref{metric:ansatz}).
For the nonsingular cases, the value of the Kretschmann invariant becomes
\begin{eqnarray}
 R_{\mu\nu\rho\sigma}R^{\mu\nu\rho\sigma}=\frac{32\,n\,{\left( n+1 \right) }^2\,{K}^{2n}}{
   \rho^{2(n+2)}}
  \left( 1+\frac{K^n}{\rho ^n}\right)^{-\frac{4(n+2)}{n}} .
\end{eqnarray}

\section{Causal Structure}

Let us consider the causal structure of these spacetime solutions.
It is enough to consider the $z_i$=constant surface with fixed angles.
Then, the metric becomes
\begin{eqnarray}
ds^2 = -H_0 dt^2+ G d\rho^2 = -H_0(dt +d\rho^*)(dt - d\rho^*).
\end{eqnarray}
Here a tortoise coordinate $\rho^*$ is defined as
\begin{eqnarray}\label{tortoise}
 \rho^* =\int^\rho \sqrt{\frac{G}{H_0}}d\rho= \int ^\rho d\rho
 \left(1+\frac{K^n}{\rho^n}\right)^{
            \frac{2}n} \,\left|\frac{1-K^n/\rho^n}{1 +
 K^n/\rho^n}\right|^{\frac{1}{n} \left( 1-\frac{\sqrt{n(n+1)}}{\sqrt{\bar
    a^2+1 -\frac{(a+1)^2}{p+n+1} }}  \right)}.
\end{eqnarray}
The ingoing and outgoing null coordinates are defined by
\begin{eqnarray}
v= t+\rho^* , \quad u= t-\rho^* ,
\end{eqnarray}
respectively.

In order to see the causal properties of the $\rho=K$ surface, let us consider whether the null rays  can escape from the surface or not.
The geodesic motion of an outgoing light in $(v,\rho)$ coordinates becomes
\begin{eqnarray}
\frac{d\rho}{dv} = \frac{1}{2}\sqrt{\frac{H_0}{G}} \sim
    |\rho-K|^{-q}
\end{eqnarray}
in the vicinity of the $\rho=K$ surface.
Here $q$ is given by
\begin{eqnarray}
q=\frac{1}{n}\left(1-\sqrt{\frac{n(n+1)}{\bar a^2+1
        -\frac{(a+1)^2}{p+n+1} }}\right) \geq -1,
\end{eqnarray}
where the equality holds only for $a_i=1/(n+1)$ for all positive $i$.
The elapsed value $v$ for a light traveling from $\rho =K+\epsilon$ to  $\rho =\rho'$ outside becomes
\begin{eqnarray}
\Delta v = 2\int_{K+\epsilon}^{\rho'}\sqrt{\frac{G}{H_0}} d\rho
    \sim \left\{\begin{tabular}{ll}
            $\displaystyle
            \ln\left|\frac{\rho'-K}{\epsilon}\right|$ & for
            $\displaystyle a_i=\frac1{n+1}, ~~\forall i, $ \\
        $\displaystyle
            \ln\left|\rho'-K\right|^{q+1 }
            -\epsilon^{q+1}$ & for
            $\displaystyle a_i\neq \frac1{n+1},~~ \exists i $. \\
            \end{tabular}\right.
\end{eqnarray}
For the case of $a_i=1/(n+1)$, which corresponds to the
Schwarzschild black brane, $\Delta v\rightarrow \infty$ as $\epsilon
\rightarrow 0$. This implies that the light cannot escape from the
$\rho=K$ surface and consequently the $\rho=K$ surface is indeed an
event horizon.
On the other hand, $\Delta v$ takes a finite value for other cases.
Therefore, the light can actually escape from the $\rho=K$ surface
in a finite time.
In this sense, the surface is not an event horizon and, in fact, is
interpreted as a naked singularity because of the curvature
singularity there.

\begin{figure}[htbp]
\begin{center}
\begin{tabular}{cc}
\includegraphics[width=.48\linewidth,origin=tl]{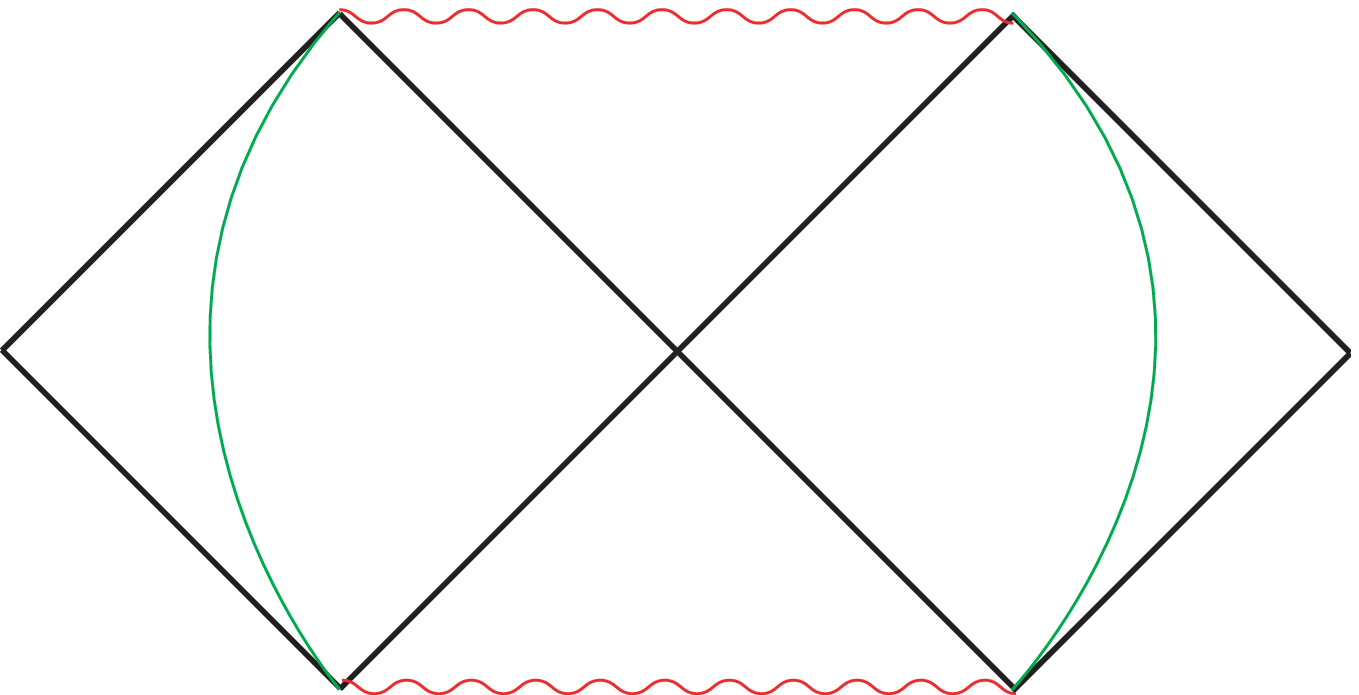}
 \put(-218,70){\small $\rho = 0$}
 \put(-194,40){\small $\rho = $ const.}
 \put(-143,78){\small $\rho = K$}
 \put(-97,78){\small $\rho = K$}
 \put(-40,40){\small $\rho = $ const.}
 \put(-18,75){\small $\rho = \infty$}
 \hspace*{1mm}
 \put(80,50){\small $\rho = K$}
 \put(150,50){\small $\rho = K$}
 &
  \hspace*{1mm}
 \includegraphics[width=.48\linewidth,origin=tl]{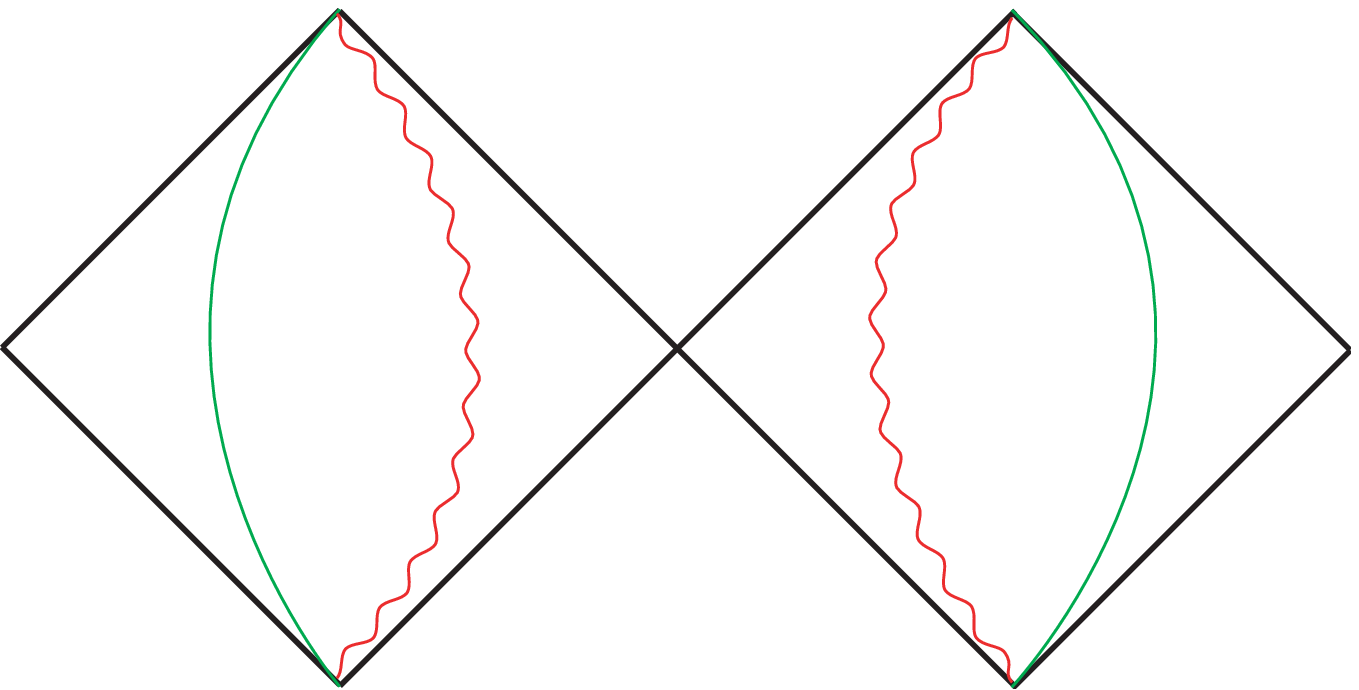}\\

 (a) & (b)\\
\end{tabular}
\end{center}
\caption{Penrose diagram at a $z_i=$constant slice. The angular
coordinates are suppressed here. Fig.~(a) is for the case that all
tension-to-mass ratio is $1/(n+1)$, which corresponds to the
Schwarzschild black brane. Fig.~(b) is for the case having naked
singularity at the $\rho = K$ surface.} \label{fig2}
\end{figure}

The Penrose diagram for the spacetime at $z_i$=constant is given in
Fig.~\ref{fig2}. Fig.~(a) is for the case that all tension-to-mass
ratio is $1/(n+1)$, which corresponds to the Schwarzschild black
brane. The surface of $\rho = K$ is indeed an event horizon. The
region of $0 < \rho < K$ is not the inside of $\rho = K$ surface,
but it is another copy of the spacetime as well known in the
isotropic coordinate system. The inside region can be obtained by
analytic continuation of the spacetime covered by $0 < \rho <
\infty$, revealing a space-like singularity at the center. The case
of Kaluza-Klein bubble is not shown here. Fig.~(b) is for the case
other than these two. Here the $\rho = K$ surface becomes a
time-like singularity ({\it i.e.}, a naked singularity), and the
spacetime cannot be analytically continued beyond this surface like
the spacetime around the space-like singularity at the center in
Fig.~(a).

\section{Conclusions}
We have investigated the geometrical properties of static vacuum
$p$-brane solutions of Einstein gravity in $n+p+3$ dimensions, which
have spherical symmetry of $S^{n+1}$ orthogonal to the
$p$-directions and are invariant under the translation along them.

The solutions are characterized by $(p+1)$ parameters, that are mass
density $M$ and $p$-number of tension densities.
Interestingly, some important geometric properties such as the area
of $S^{(n+1)}$ and the total spatial volume~(\ref{r}) are
characterized only by the three parameters ($M, a, \bar a$), not by
the individual value of tension ratios $a_i$, where $a$ and $\bar
a^2$ are the summations of tension to mass ratio and of their
squares, respectively, as defined in Eq.~(\ref{a}).
The surface area of $S^{n+1}$ sphere shows interesting behavior
depending on the values of $a$ and $\bar a$. There are three
classes. Namely,
i) $ \frac{(n+1)^2+p}{(p+n+1)^2} (a+1)^2-\bar a^2 >
1$;
   As $\rho$ decreases, this area does not decreases monotonically, but
increases again as $\rho$ approaches to $K$. In other words, the
spatial geometry of the $z_i=$ constant surface behaves like a
wormhole geometry. ii) $ \frac{(n+1)^2+p}{(p+n+1)^2} (a+1)^2-\bar
a^2 = 1$;
   This case is designated by the thick curve ($A_{n+1}$)  in Fig. 1. The Schwarzschild black $p$-brane and the Kaluza-Klein bubble solutions belong to this class.
  The area monotonically decreases to a non-vanishing finite value at $\rho=K$.
iii) $ \frac{(n+1)^2+p}{(p+n+1)^2} (a+1)^2-\bar a^2 < 1$;
 The area monotonically decreases to zero at $\rho = K$.

The spatial volume $V_p$ of unit extra $p$-directions takes a non-vanishing finite value only when $a= p/(n+1)$. As $\rho$ decreases to $K$, it monotonically increases to infinity for $a<p/(n+1)$ and monotonically decreases to zero for $a>p/(n+1)$.
Although each $A_{n+1}$ and $V_p$ behaves variously as $\rho$ goes to $K$, the total volume $A=A_{n+1}V_p$ always shrinks to zero except for the case of $a_i=1/(n+1)$ for all $i$.
It is interesting to see that this value is independent of the number of extra dimensions $p$.
For such exceptional case, $A$ takes a finite value and it corresponds to the Schwarzschild $p$-brane solution.
The curvature square diverges at $\rho=K$ except for the cases of the Schwarzschild black $p$-brane~(\ref{Schwarz}) and Kaluza-Klein bubble~(\ref{bubble1}).
This curvature singularity at the surface $\rho=K$ turned out to be naked.

Some geometrical properties explicitly depend on each value of
tensions. In particular, the spatial geometry for each $z_i$
directions are dependent on $a_i$ in addition to $a$ and $\bar a$.
The proper length along $z_i$ becomes neither shrinking nor
expanding as $\rho$ approaches to $K$ if the $i$-th tension takes a
specific value which is determined by the values of other tensions
as in Eq.~(\ref{ai:regular}). The regular spatial geometry in all
extra $p$-dimensions happens only for the Schwarzschild black
$p$-brane and Kaluza-Klein bubble solutions.

We have seen that the causal structure of the higher dimensional
solutions are essentially the same as those of the five dimensional
solutions~\cite{kang}. Namely, only the solutions where the value of
tensions are equal to the mass divided by the number of angular
coordinates {\it i.e.} $M/(n+1)$ have an event horizon located at
$\rho=K$. As well known, for this Schwarzschild black $p$-brane
solutions, the spacetime can be continued beyond the $\rho=K$
surface and a spacelike curvature singularity appears inside the
event horizon as shown in Fig.~2. The Kalua-Klein bubble solution,
where one of its tensions is given by mass times the number of
angular coordinates and all other tensions are equal to the mass, do
not possess an event horizon and the signature of this spacetime
changes across the $\rho=K$ surface.~\footnote{In order to see this
we need to analytically continue the spacetime region covered by $K
< \rho < \infty$ beyond the $\rho =K$ surface. Note that, in the
isotropic coordinate system, $\rho < K$ does not describe the inside
of the $\rho = K$ surface.} Other than these two cases, the
curvature singularity appears at $\rho=K$ surface and light can
escape from this surface toward the spatial infinity. Consequently,
this singularity is naked in nature.

Investigation of such singular spacetimes might be meaningless
physically. However, studying the geometrical structure of such
singular spacetimes in detail may be very important for the
following reason. It is interesting to see that the Schwarzschild
black brane is so special for geometrical properties in the whole
solution space. This fact was already pointed out in
Refs.~\cite{Bro2,stability}. In particular, when one considers
spherically symmetric perturbations with translational symmetries
along extra-directions untouched, the Schwarzschild black brane
solution is singled out because it is stable under such
perturbations whereas all other solutions possessing naked
singularities are catastrophically
unstable~\cite{stability,Tomimatsu}.~\footnote{Note, however, that
even the Schwarzschild black brane solution becomes unstable as some
perturbations depending on the extra-dimensional coordinates are
turned on, the so-called Gregory-Laflamme instability~\cite{after}.}

On the other hand, the spacetimes we considered may be formed from
the higher dimensional trans-spherical distribution of normal
matter. Our study shows that a slight deviation of tension and mass
densities from those evolving to the Schwarzschild black brane will
end up with a spacetime configuration possessing a naked
singularity. Assuming that the cosmic censorship conjecture is valid
in higher dimensions as well, we do not expect the development of a
naked singularity. This observation may indicate that some quantum
effect becomes important right before forming a naked singularity
during the gravitational collapse. In fact, Yoshimura and
Tanaka~\cite{Yoshimura} showed numerically that the naked
singularity existing at some static spacetimes near the
Schwarzschild one in parameter space becomes null if the
Gauss-Bonnet gravity term is added in 5-dimensions. It is
interesting to see how much various quantum effects change the
formation of naked singularity in trans-spherical gravitational
collapse. Finally, it is also interesting to study how much the
geometrical properties are modified if angular momentum or charge
are added in the consideration. These interesting issues deserve
future work.

\begin{acknowledgments}
This work was supported by the Korea Research Foundation Grant
funded by the Korean Government(MOEHRD, Basic Research Promotion
Fund)(KRF-2008-314-C00063) and the Topical Research Program of APCTP
and in part by the e-Science Project of KISTI. GK was supported in
part by the Korea Science and Engineering Foundation (KOSEF) grant
funded by the Korea government (MEST) (No. R01-2008-000-20866-0).
\end{acknowledgments} \vspace{1cm}


\vspace{4cm}

\end{document}